\documentclass{article}
\newcommand{\be}{\begin{equation}}
\newcommand{\ee}{\end{equation}}

\begin{document}
\large \baselineskip=14pt
\begin{center}
 {\Large\bf Determination of the cosmological
parameters from the Earth-Moon system evolution} \vspace{2cm} \\
{\large\bf A. I. Arbab}\footnote{E-mails: arbab@ictp.trieste.it,\
arbab64@yahoo.com}
\vspace{1cm}
\\
ICTP, P.O. Box 586, Trieste 34100, Italy\\
\end{center}
\vspace{2cm}
{\bf Abstract.}
We have obtained empirical laws for the variation of the Earth
parameters with geologic time that are in agreement with coral
fossil data obtained by Wells and Runcorn. Our model predicts that
the day is lengthening at a rate of $2 \ \rm ms/century$ at the
present time. The length of the day when the Earth was formed is
found to be 6 hours and the synodic month to be $\rm 56 $ days.
The angular momentum of the Earth-Moon system is found to be
increasing  with time. The origin of the presently observed
acceleration of the Moon is explained. It is predicted that the
Moon is approaching the Earth at a present rate of $\rm 1.5\
cm/year$ and with an acceleration of $20.6\ \rm arc\ sec/cy^2.$ It
is  shown that the Moon has never been close to the Earth in last
$\rm 4.5$\ million years.
\vspace{1cm}
\\
{\bf 1. Introduction}
\\
\\
The Einstein General Theory of Relativity (GTR) is one of the most
fascinating theories that people have come to know. Einstein
himself, after the overwhelming success of his theory when applied
to the Sun, constructed a model for the whole universe. The
Einstein model was a static model, in which stars (or galaxies) do
not move, as  appeared to him at that time. After the universe was
found to be expanding with time, Einstein abandoned his model.
Obtaining a static model of collection of matter alone is not
possible, and this led Einstein to hypothesize a repulsive force
(later known as the cosmological constant ($\Lambda$)) to hold his
universe from collapse. Einstein regretted the addition of this
constant when it appeared to him that the expansion plays the role
of repulsion ($\Lambda$). Later, Einstein and de Sitter
constructed a model for the universe(now known as the Standard
Model) in which matter and radiation are distributed homogeneously
and isotropically. That model is successful in many regards: it
predicts an existence of a cosmic background radiation and a
primordial synthesis of helium and hydrogen in the earliest stage
of expansion of the universe. Such predictions are now well
established facts about the present universe. However, the {\it
Standard Model} is fraught with other problems which could not be
solved in the framework of the Einstein-de Sitter model. An idea
that the universe had once expended at an enormous rate was
brought by Guth [1980] and  Linde [1981] which is known as the
{\it Inflationary Scenario}. This scenario solves, most if not
all, of the standard model puzzles. The {\it Standard Model} gives
an age for the universe of $t=\frac{2}{3}H^{-1}$, where $H$ is the
Hubble constant. This value is smaller than those obtained by
astronomers. Cosmologists have found that if one still retains the
concept of the cosmological constant the age problem can be
resolved. However, a pure constant might pose a problem, from the
point of view of particle physicists who estimated its value at
the Planck time to be  120 orders of magnitude smaller than its
present one.

In fact, the idea of a variable is associated with Dirac [1937]
who noticed some puzzling coincidence between atomic and cosmic
scales. He found the age of the universe in terms of atomic time
to be about $10^{40}$, the ratio between the electric and
gravitational forces of an electron and a proton (e.g., in
hydrogen atom) to be about $10^{40}$ and the ratio of the Hubble
radius (observable radius of the universe) to the classical radius
of electron to be about $10^{40}$. He thought that the coincidence
of these dimensionless huge numbers is not accidental, but must
have a physical origin. In order that this coincidence to hold at
all times, and not just for the present time, he proposed that the
gravitational constant has to change with time inversely, i.e.,
$G\propto t^{-1}$. Therefore, to Dirac, the expansion of the
universe lead to a weakening of the gravitational constant. Later,
Brans and Dicke [1961] (BD) formulated a field theoretic model in
which $G$ is related to a scalar field $(\phi)$, that shares the
long range interaction with gravity, viz. $G\propto \phi^{-1}$.
Their theory predicts a decreasing $G$, as in Dirac model,  but
with a lesser rate. For conformity with the present observations
regarding the solar system, BD theory becomes indistinguishable
from GTR. The BD theory is set to satisfy Mach principle which
asserts that our local physics is affected by the presence of
distant matter in the universe. \\ The variation of $G$ with time
would have numerous geological and astronomical consequences. For
instance, the luminosity of the Sun depends on $G$. The distance
between the Earth and the Moon, the length of the month, the
Earth's surface temperature, the length of the day and the Earth's
radius would all be affected. Shapiro {\it et al} [1971] have set
an upper limit to the present variation of $G$, viz., $|\frac{\dot
G}{G}|_0\le 4\times 10^{-10}y^{-1}$. A comprehensive investigation
of the effect of the variation of $G$ can be found in Wesson
[1978]. \\ The idea that gravity increases with time was proposed
by several authors [Abdel Rahman, 1990, and references therein].
Canuto [1976] suggested a variation of the form $G\propto t $. In
line with that, we have recently proposed a cosmological model and
have found that the gravitational constant has to increase in
order to resolve the cosmological contradictions of the {\it
Standard Model} with observations. This represents a minimal
change of the {\it Standard Model} to fit the current
observational data. We have found that $G\propto t^2$ in the early
universe, but generally evolves as $G\propto t^{(2n-1)/(1-n)}$,
where $n$ is related to viscosity of the cosmic fluid. We have
shown in that work that many non-viscous models are equivalent to
viscous models with variable gravity [Arbab, 1997]. Our model does
not determine $n$ exactly but rather imposes a constraint that
$\frac{1}{2}\le n\le 1$ in order to comply with the present
observational data. All of our cosmological quantities depend only
on the parameter $n$. Very recently, a similar variation of $G$
(without viscosity) in terms of another parameter $\beta$, whose
value determines the whole cosmology is also considered [Arbab,
1999]. The universe is consequently shown to be accelerating at
the present time. This acceleration requires $n>\frac{2}{3}$ (or
$\beta >3$) [Arbab, 1999].

The purpose of this paper is to determine $n$ from a different
source of data that was not tried before. We remark that the
evolution of the universe affects the Earth-Moon system
indirectly, by allowing the gravitational constant to change with
time at the present epoch. A fixed value for $n$ (or $\beta$) is
found  from studying the evolution of the Earth-Moon system, viz.
$n=0.7$ (or $\beta=12$). This immediately implies that the Moon
must be accelerating in its orbit while approaching the Earth. The
same effect should be observed in the motion of the Earth around
the Sun.
\\
\\
{\bf 2. The Sun-Earth-Moon System}
 \\
\\
  Kepler's second and third laws governing the Sun-Earth-Moon system can be 
written as
  \be
  G^2[(M+m)^2m^3]T=2\pi L^3, \ \ \ \ \ \ \ G[(M+m)m^2]r=L^2,
  \ee
  \be
G^2[(M+M_s)^2M^3]Y=2\pi N^3, \ \ \ \ G[(M+M_s)M^2]R=N^2,
 \ee
where $m$= mass of the Moon, $M$= mass of Earth, $M_s$= mass of
the Sun,  $L$= the orbital  angular momentum of the Moon, $N$= the
orbital angular momentum of the Earth, $r$= the Earth-Moon
distance, $R$= Earth-Sun distance, $Y$=the number of days in a
year and $T$= the sidereal month.
\\
It is believed that $N$ has remained constant throughout the Earth
history while $L$ changes with time. The synodic month, $T_{sy}$,
is related to the sidereal month by the relation
\be
T_{sy}=\frac{T}{(1-\frac{T}{Y})}
 \ee
From the above equations one can write \be
\frac{Y}{Y_0}\frac{T_0}{T}=(\frac{L_0}{L})^3, \ee (hereafter, the
subscript `0' refers to the value of the quantity at the present
time).\\ Based on the slowing down of the Earth's spin with time,
the conservation of total angular momentum of the Earth-Moon
system dictates that the Moon angular momentum has to increase. It
is evident from eq.(1) that if $G$ increases then so will $L$. It
can also be inferred from eq.(1) that if $L$ increases then $r$
must increase. In this case the Moon will lose its kinetic energy
as it goes away from the Earth. However, observations show that
the Moon is accelerating in its orbit [Dickey 1994], which, in
turn, means that the Moon is gaining kinetic energy. Here lies the
conflict and this is why this acceleration is called anomalous
acceleration. In the present approach the variation of the Earth
parameters with time is related to that of $G$. For instance, the
acceleration of the Moon is a consequence of the increase of $L$.
This would require the distance $r$ to decreases in such a way
that $L$ increases. Therefore, our model predicts, rather than
conflicts with, the presently observed acceleration of the Moon.
Our prediction is well within the observational limits. Moreover,
as a very close approach of the Moon in the past did not occur,
this model resolves the problem of the short age of the Moon.
\\
\\
{\bf 3. Coral and Bivalve Fossil data}
\\
\\
Data about the past rotation of the Earth was obtained from fossil
coral by Wells [1963], who concluded that the day was shorter in
the past. In his  systematic study of coral fossils, Wells was
interestingly able to relate the number of ridges recorded by the
corals to the number of days in a year, during the time when that
corals had lived. He could go back with this analysis up to
$\rm600$\ million years ago. Investigations  that employed coral
fossils and  related to the synodic month were done by Panella
[1968], Munk [1966] and Scrutton [1964]. Barry and Barker [1975]
employed bivalve fossil data and obtained a similar conclusion
about the length of month in the past times. It has thus become
clear that these corals can be used as chronometer, where changes
in a length of day can be noticed. This method is clearly
advantegous to radiometric dating, where only a time passage of
million years can be differentiated. Wells's data is shown in
Table 1 and our corresponding data is shown in Table 2. Using
Wells and Scrutton data, Runcorn [1962, 1964] concluded that the
ratio of the angular momentum of the Moon 400 million years ago,
to the present one ($L_0$) to be \be \frac{L_0}{L}=1.016\pm 0.003
\ .\ee
\\
{\bf 4. The Earth-Moon system and the cosmological parameters}
\\
\\
Using the results of Wells and Runcorn, one finds the variation of
$G$ with time to be as follows \be G=G_0(\frac{t}{t_0})^{1.3} \
,\ee where $G_0$ is the present value of Newton's constant and
$t_0$  the present age of the universe. Wells and Runcorn data
require that the age of the universe to be $11\times 10^9$ years.
Hence, our model predicts that \be |\frac{\dot G}{G}|_0=1.1\times
10^{-10}\rm y^{-1}\ . \ee This age is found to be in agreement
with the present astrophysical estimates. However, ages obtained
from astronomical data usually involve some uncertainties ($\rm
10-15$\ billion years). This constraint on the age of the universe
implies that the universe is accelerating at present with an
acceleration parameter of $-0.1$. This means that the contribution
of ordinary matter to the total energy density in the universe is
$60\%$.

Equation (6) can be viewed as representing an effective value for
the gravitational constant ($\rm G_{eff.}$). Thus all
gravitational interactions couple to matter with this effective
rather than `bare' constant $(\rm G_0=6.67\time
10^{-11}Nm^2kg^{-2}$). The introduction of the effective
gravitational constant should leave our physical laws intact (form
invariant). In effect, all old quantities have to be replaced with
their corresponding effective ones (see below). With this
prescription one can employ the old equations of Kepler, Newton,
etc. for describing the present phenomena without any further
complications. \\ Equation (6) can be written as follows \be
G_{\rm eff.}=G_0 f(t), \ee with $f(t)=1$, for $t=t_0$. The idea of
dealing with an effective quantity rather than a `bare' one has
already been applied for the motion of the electron in a solid,
where it was shown that the electron moves with a different
(effective) mass that is related to its `bare' one.

In the present study, this anstaz will have lot of implications on
the motion of celestial objects. One of the immediate consequences
of this assertion for Newton's law of gravitation is that a system
in equilibrium tends to exhibit an apparent acceleration towards
its center. The other physical (effective) quantities can be
expressed only in terms of their present values and the past time
(measured from the present time epoch backward). The number of
days in a year and the length of the day are given by \be
Y=Y_0(\frac{t_0}{t_0-t})^{2.6},\ \ \ \ \  \,
D=D_0(\frac{t_0-t}{t_0})^{2.6} \ee In fact, the time duration of
the year has not changed, but the number of days in a year changes
because the day is getting longer  as time goes on due to tidal
effects of the Moon on the Earth. This means that in the past, the
number of days in the year was bigger than now.

Recently, McNamara and Awramik [1992] have concluded, from the
study of {\it Stromatolites}, that at about 700 million years ago
the number of days in a year was $\rm 435$\ days and the length of
the day was $\rm 20.1$\ hours. In fact, our model predicts this
value to correspond more accurately to $\rm 715$\ million years
ago. This indicates that this approach is reliable and can be used
safely to get some clues about the nature of life that was
prevailing at a particular time in the past. In this manner, one
can use these {\it Stromalotites} as chronometer for dating
geological rocks in which they are found.

Sonett {\it et al} [1996] have used {\it laminated tidal
sediments} to determine the number of days in a year some $\rm
900$ \ million years ago and inferred from these records that the
number of days in a year was $\rm 456$\ days, and that the length
of day was $\rm 19.2$\ hours, considering only the lunar
component. They also concluded that the sidereal month to be $\rm
23.8$\ days (present epoch day). This is in a agreement with the
value we obtain for this period (see Table 2). In fact, {\it
Stromatolites} are believed to be dated back to the earliest stage
of the Earth formation ($\rm 3.5$\ billion years ago-see McNamara
and Awramik).

Hannu [1991] made a statistical formula from the data of the
corals and bivalve and found that $\rm 3.556$\ billion years ago
the year contained $\rm 1009$\ days. All these results are in good
agreement with our model predictions.

Our model predicts that when the Earth was formed the length of
the day was $\rm 6$\ hours, the year having about $\rm 1434$\ days
and the synodic month being about $\rm 56$\ days. This is because
the Moon orbited the Earth with rather a smaller velocity and at a
larger orbit. Studies indicate that billion of years ago, the
length of the day was between $\rm 5$\ and $\rm 6$\ hours. These
results have been confirmed by an unexpected finding [William,
1994]. Moreover, Schopf [1983] has found the length of the
day 4.5 billion years ago to be 5-7 hours. \\
As it difficult to find fossils that dated back to $\rm 4.5$\
billion years ago, the information about the primitive Earth can
only be extrapolated from these data. For this reason our model
becomes a very useful tool for those who are keen to know about
the early evolution of the Earth-Moon system.
\\
\\
{\bf 5. The Moon's Acceleration}
\\
\\
The orbital angular momentum ($L$), the length of the sidereal
month ($T$), the Moon orbital velocity ($v$), and the Earth-Moon
distance ($r$) are give by \be L=L_0(\frac{t_0-t}{t_0})^{0.44}\ ,
\ \ \ \ T=T_0(\frac{t_0}{t_0-t})^{1.3}, \ee \be
r=r_0(\frac{t_0}{t_0-t})^{0.44}\ ,\ \ \ \ \
v=v_0(\frac{t_0-t}{t_0})^{0.88}. \ee It is shown by Dicke [1962]
that the Earth radius ($d$) is related to $G$ as $d\propto
G^{-0.1}$, hence \be d=d_0(\frac{t_0}{t_0-t})^{0.13}. \ee The
lunar and Earth angular velocities are given by \be
\omega=\omega_0(\frac{t_0}{t_0-t})^{2.6}\ \ , \ \
n=n_0(\frac{t_0-t}{t_0})^{1.3}\ , \ee where
$\omega=\frac{2\pi}{D}$, $n=\frac{2\pi}{T}$, $n_0=2.661699\times
10^{-6}$ rad/sec, $\omega_0=7.2921\times 10^{-5}$ rad/sec.\\ It is
evident that an increasing gravitational constant leads to a
contraction of the Earth radius. Such a variation can be tested
with observation. In fact, the idea of a contracting Earth was
once favored by geologists, as it helps interpret the process of
mountain building. Bursa [1987] has shown that the resonant
condition between the Earth and Moon is such that the length of
the month is $\rm 50.7$\ days, a distance of $\rm 581098$\ km,
which corresponds to a time of 3.2 billion years ago. According to
our model the Earth had once a similar condition, viz., at a time
of $\rm 4.16$\ billion years ago when the  Earth-Moon distance was
$\rm 473774$\ km. If that was the case, our model may provide the
answer to the question of the origin of the Moon that preoccupied
scientists for several decades.
\\ According to our model the Earth is contracting at a present
rate of $\rm 0.1 mm/year$. A contraction in the Earth will result
in an apparent increase in the sea-covering water areas.

Equation (12) implies that the Moon is approaching the Earth at a
present rate of $\rm 1.5\ cm/y$, while Apollo mission to the Moon
suggests a recession of the Moon of $\rm 3.8\ cm/y$ [Dickey {\it
et al}, 1986]. This result is the only disagreement that our model
has with the present  observational data. The present recession
obtained from the Apollo mission when extrapolated backward (to
$\rm 1.4-2.3$\ billion years) would imply that the Moon was almost
touching the Earth's surface. However, the age of Earth-Moon
system is believed to be more than this time (about $\rm 4.5$\
billion years ago). This is a real difficulty to Apollo data
regarding the origin of the Moon. However, there is so far no
geological evidence for such a situation. \\ Slichter [1963]
remarked that`` if for some unknown reason the tidal torque was
much less in the past than in the present (where `present' means
roughly the last $\rm 100$\ million years), this would solve the
problem ". He concluded that `` the time scale of the Earth-Moon
system still presents a major problem." The Moon could not remain
an intact body much closer to the Earth than about $\rm 18000$\
km. Brosche [1984] remarked that ``from the results of paleotides,
it seems probable that the Moon was not in a `narrow' state,
especially if there were no oceans during the first billion
years."
\\ In fact, the increasing angular momentum does not
necessarily imply that the Moon is receding (as immediately taken
by many people). We could resolve this situation even with a Moon
approaching the Earth (a decreasing $r$). That is because the Moon
is accelerating in its orbit and this acceleration takes into
account the decrease of the Earth-Moon distance. This is evident
from our eq.(11).

While our model predicts a lunar acceleration of ($\rm 20.6\ arc\
sec /cy^2$), the present estimate of lunar acceleration  from
lunar laser ranging amounts to ($-24.9\pm 1.0)\ \rm\ arc\
sec/cy^2$ [Dickey {\it et al}, 1986]. Christodulidis {\it et al}
[1988] have found the total contribution from tidal effects to be
$-21.4 \rm\ arc\ sec/cy^2 $. This result is very close to our
prediction. Spencer and Jones [1939] found the acceleration to be
$ -22\pm 1.1 \rm \ arc\ sec/cy^2$. Lambeck [1980] has shown that
the rate of change of the lunar and Earth angular momenta to be
related by \be \dot \omega=(51\pm 4)\  \dot n\ .\ee Lambeck [1980]
and Stacey [1977] argued that `` tidal dissipation must have been
lower in the past ". Equation (13) gives the following \be \dot
\omega=-54.8\ \dot n\ .\ee This represents another agreement with
observation. Equation (13) implies \be \dot\omega=-5.47\times
10^{-22}\rm radian/sec^2 \ .\ee where its shown by Christodulidis
[1988] to give \be \dot\omega=(-5.98\pm \rm\ 0.22)\times
10^{-22}\rm radian/sec^2 \ .\ee which shows how our prediction is
closed to the observed one. In fact, Lambeck [1980] has found
$\dot\omega=-5.48\times \rm 10^{-22}rads^{-2}$ to be his consensus
value from astronomical observations. Murray [1957] has shown that
the deceleration derived from ancient lunar eclipse is in an
excellent agreement with that deduced from Hipparchus's
observations. He combined the two independent determinations and
adopted that the Moon acceleration to be $(\rm -21\pm 3)arc\
sec/cy^2$. Calame and Mulholland [1978] considered the possibility
of the time variation of $G$ in inducing any tidal acceleration.
They pointed out that $ \dot n $ is supposed to be constant in
nearly all discussion pertaining to it, as no geophysically
plausible mechanism can account for a significant variation over
historic time. They observed that all of the various solutions lay
in the range $\rm -22$\ to $-26\ \ \rm arc\ sec /cy^2$. They
concluded that ``it will require several more years before the
cosmological question of $\dot G$ can be resolved with any
confidence". Calame and Mulholland [see Brosche and Sundermann
(1978)], when correcting Williams [1974] data for the origin of
longitude, obtained an acceleration of the Moon of $-20.7\pm
1.6\rm \ arc\ sec/cy^2$.
\\
It is  speculated that as the  Moon recedes far away from the
Earth the tidal effects become unimportant and the length of the
month and that of the day will be both equal to about $55$ days
(present epoch days). We have found that the month in the
beginning was $56$ days and the Moon was in its farthest point
from the Earth. This picture on one hand resembles the situation
that our model predicts.
\\
One of the competing scenarios about the origin of the Moon is the
fission theory by Darwin. This requires the Earth to have rotated
at a period of $\rm 2.5$\ hours in the past. As this was not the
case, our model  rules out this scenario.
\\ The rapid rotation of the Earth in the beginning led to tremendous
geologic effects on the Earth's motion. It gave rise to various
geologic activities: volcanos, earthquakes, mountain building,...,
etc. Since the Earth adjusts itself so as to be in equilibrium,
the slow-down of Earth rotation would have its influences on the
evolution of the Earth-Moon system. The continental drift
suggested by geologists would have its causes in the ever-changing
rotational motion of the Earth and the tidal forces exerted by the
Moon. The rapid rotation of the Earth in the past must have
affected our past climate.
\\
We notice from Fig.1 that the number of days in a year  decreased
uniformly in the beginning but curved during some time and again
decreases steadily. The time when this curvature occurred could be
due to a significant emergence of water on the Earth. This is
because as water became abundant, tidal forces would have slowed
the Earth rotation at a bigger rate. However, this period, which
lies between $\rm 1.1 - 3.8$\ billion years ago, is the time that
life was first appeared on the Earth as suggested by scientists.
Researchers in oceanography have to set the proper limits for this
an outstanding event in the Earth past history.
\\
The escape velocity and the thermal velocity  of a gas molecule
are related by $v_{es.}=\sqrt{\frac{2GM}{R}}$,
$v^{th.}=\sqrt{\frac{3kT}{m}}$, where $k$ is the Boltzman
constant. For the gas molecules to immigrate from the Earth
gravity it must have $v^{th.}>>v^{es.}$. The luminosity
($\cal{L}$) of the Sun is related to the Earth's temperature by
the black-body formula, $T^4=\frac{\cal{L}}{d^2}$. But
${\cal{L}}\propto G^8$ [Weinberg, 1972]. Therefore, the evolution
of the escape velocity and the temperature of the Earth can be
written as \be \theta=\theta_0(\frac{t_0}{t_0-t})^{3.25}, \ \ \  \
\ v^{es.}=v^{es.}_0(\frac{t_0}{t_0-t})^{1.56}\ . \ee The escape
velocity for the gases making the atmosphere was not enough to
leave the Earth since the Earth started cold at a temperature of
$\rm 54$\ K. This is why a cold origin of the Earth is favored
over the hot one. Hence the Earth, most likely, has retained its
primordial atmosphere. A  cold start of the Earth would have
facilitated the development of life on the primitive Earth. The
appropriate conditions prevailing at that time suggest that life
could have originated inside  the sea, where the temperature was
higher, originating from rapid frictional motion between Earth's
plates that provided sufficient heat inside the sea. And when the
outside conditions have changed life gradually moved to the
surface, where light was abundant to trigger the essentials of
life.
\\
\\
{\bf 6. The Future of the Earth-Moon System}
\\
\\
As the Earth gradually heats up, the ice caps on the poles will
start to melt. This would increase the tidal effects on the Earth
while the Moon approaches it. Thus, in the distant future the
tidal effects will become very strong and the Moon will enter a
region of destablization (Roche limit). As the tidal forces
increase the rotational velocity of the Earth will decrease until
it reaches a stage when the Earth can no longer remain in
equilibrium. This will lead to big disturbances in the Earth's
interior allowing the molten hot core material to come to the
surface. This disruption will result in many geologic activities,
such as,  volcanoes, earthquakes, etc. The Earth's temperature
will become higher and life will again be pushed back to the sea,
as it once started in!
\\
\\ {\bf 7. Conclusion}\\
\\
We have developed a model that describes the evolution of the
Earth-Moon system from its inception to the present. This system,
which dates back to $\rm 4.5$\ billion years, had encountered
various conditions that finally shaped the present Earth-Moon
system as we observe today. The history of this system is not
lost, but embedded in different forms, in biological samples that
had punctuated clocks and the rocks that embodied and preserved
them. Our important prediction is that the Moon is approaching the
Earth at a rate of $\rm 1.5cm/y$ thus causing the Moon to
accelerate. The consequences of this model may have important
impacts on geology, astronomy, theology, religion, arts,
geography, biology, archeology, etc. The full data pertaining to
the Earth in the past is tabulated and the necessary graphs are
plotted. Data from paleontology and geology can not go back to the
first instances of the formation the Earth. As there is no rock
available more that $\rm 3.8$\ billion years ago. This justifies
the necessity of having a model that can extend these data. For
these issues we have to recourse to theories about the formation
of the celestial objects to know the initial parameters describing
our Earth-Moon system and other similar system. We have developed
a complete and consistent simple mathematical method for handling
the tidal dissipation, that agrees well with observations. The
present model can be applied to other coupled systems in the solar
system. The expansion of the universe is found to affects the
evolution of the Earth-Moon system by imparting a small change in
the gravitational constant. Further investigations are going on to
ravel other hidden subtleties.
\\
\\
{\bf Acknowledgements}
\\
\\
I would like to express my thanks and appreciation to the Abdus
Salam ICTP  for hospitality, the Swedish Development Cooperation
Agency (SIDA) and the Associate Scheme and Federation for
financial support. I also wish to thank Dr. H. Widatallah for
stimulating discussion.
\\
\\
{\bf References}
\\
\\ Abdel Rahman, A.M.-M., {\it Gen. Rel. Gravit.22}, 655, 1990\\ Arbab, 
A.I. {\it Gen.Rel. Gravit. 29}, 61,
1997\\ Arbab, A.I., http://xxx.lanl.gov/abs/gr-qc/9905066\\ Barry,
W.B.N., Baker, R.M. (in Rosenberg, G.D., and Runcorn: {\it Growth
rhythms and the history of the Earth's rotation}, Willey
Interscience, New York, 1975)\\ Brans, C, Dick, R.H., {\it Phys.
Rev.124}, 1961\\ Brosche, P., {\it Phil. Trans. R. Soc. Lond.
A313}, 71, 1984\\ Bursa, M., {\it Bull. Astron. Inst. Czechosl.
V.38}, 321,1987\\ Calame, O. and  J.D. Mulholland, {\it Science
199}, 977,1978\\ Canuto, V et al, {\it Nature 261}, 438, 1976\\
Christodulidis, D.C., and Smith, D.E., {\it J. Geophys. Res. V93}
6216, 1988\\ Dicke, R.H., {\it Sience, 138}, 653, 1962\\ Dickey,
J.O. et al. {\it Science 265}, 482, 1994\\ Dirac, A.P.M., {\it
Nature 139}, 323, 1937\\ Frank, D.,  {\it Physics of the Earth},
John Wiley \& Sons,1977\\ Guth, A., {\it Phy. Rev. D23}, 347,
1981\\ Jeffreys, H, The Earth: {\it Its origin and physical
constitution}, 1976\\  Lambeck, Kurt, {\it The Earth's Variable
Rotation - Geophysical causes and consequences}, Cambridge
University Press, 1980\\ Linde, A., {\it Phys. Lett.B108}, 389,
1982\\ McNamara, K.J, Awramik, S.M., {\it Sci. Progress 76}, 345,
1992\\ Munk, WJ, {\it The Earth-Moon system}, New York, Plenum,
1966\\ Murray, C.A., Mont.Not. Roy. Astr.Soc. 117, 478, 1957\\
Pannela G., et al. {\it Science 162}, 792, 1968\\ Runcorn, S.K.,
{\it Nature 204}, 823, 1964\\ Runcorn, S.K., {\it Nature 195},
1248, 1962\\ Runcorn, S.K., Nature 193, 311, 1962\\ Runcorn, S.K.,
{\it Paleogeophysics}, Academic Press, London, 1970\\ Schopf,
J.William, {\it Earth Earliest Biosphere: Its origin and
Evolution}, 1983, Princeton University Press, New Jersey, U.S.A.\\
Scrutton, C.T., {\it Paleontology. 7}, 552, 1964\\ Shapiro, I.I.,
et al,  {\it Phys. Rev. Lett. 26}, 27, 1971\\ Slichter, Louis B.
{\it J. Geophys. Res. 68}, 14,1963\\ Sonett et al, {\it Science
273}, 100, 1996\\ Spencer and Jones (see Brosche, P., and
Sundermann, J., {\it Tidal friction and the Earth rotation},
Springer-Verlag, Berlin 1978)\\ Weinberg, S, {\it Gravitation and
Cosmology}, Wiley \& Sons, 1972\\ Wells, J., {\it Nature 197},
948, 1963\\ Wesson, P.S, {\it Cosmology and Geophysics}, Oxford
University Press 1978\\ William K. Hartmann and Chris Impey,
Wadsworth, Inc. HTML by Guy K. McArthur, 1994\\ Williams, J.G,
private communication, 1974 (see Brosche and Sundermann, pp.43,
1978)\\
\newpage
\begin{table}
\caption{Data obtained  from {\it fossil corals and radiometric
time}}\vspace{0.4cm}
\begin{tabular}{|r|r|r|r|r|r|r|r|r|r|}
\hline Time*  & 65 & 136 & 180 & 230 & 280 & 345 & 405 & 500 &
600\\ \hline solar days/year & 371 & 377 & 381 & 385 & 390 & 396 &
402 & 412 & 424\\ \hline
\end{tabular}
\end{table}
\begin{table}
\caption{Data obtained from the {\it principle of increasing
gravity}}\vspace{0.4cm}
\begin{tabular}{|r|r|r|r|r|r|r|r|r|r|r|} \hline Time*
& 65 & 136 & 180  & 230 & 280  & 345  & 405 & 500 & 600\\ \hline
solar days/synodic month
 & 29.74 & 29.97 & 30.12 & 30.28 & 30.45 & 30.78 & 30.89
&31.22 & 31.58\\ \hline solar days/sidereal month  & 27.53 & 27.77
& 27.91 & 28.08 & 28.25 & 28.48 & 28.69 & 29.02 & 29.39\\ \hline
synodic month/year  & 12.47 & 12.59 & 12.66 & 12.74 & 12.82 &
12.93 & 13.04 & 13.20 &  13.38\\ \hline sidereal month/year  &
13.47 & 13.59 & 13.66 &13.74 & 13.82 & 13.93 & 14.04 & 14.20 &
14.38\\ \hline solar days/year  & 370.9 & 377.2 & 381.2 & 385.9 &
390.6 & 396.8 & 402.6 & 412.2 &   422.6\\ \hline length of solar
day (hr) & 23.6 & 23.2 & 23.0 & 22.7 & 22.4 & 22.1 & 21.7 & 21.3 &
20.7\\ \hline
\end{tabular}
\end{table}
\begin{table}
\begin{tabular}{|r|r|r|r|r|r|r|r|r|r|r|r|}
\hline Time*   & 715 & 900 &1000 & 1200 & 1400 & 2000 & 3000 &
3500 & 4500\\ \hline solar days/synodic month  & 32.00 & 32.72 &
33.11 & 33.93 & 34.79 & 37.63 & 43.48 & 47.09 & 56.26\\ \hline
solar days/sidereal month &
  29.81 & 30.53 & 30.92 & 31.75 & 32.61 & 35.46 & 41.33 &
44.90 & 54.14 \\ \hline synodic month/year  & 13.59 & 13.94 &
14.13 & 14.54 & 14.96 & 16.35 & 19.23 & 20.99 & 25.49\\ \hline
sidereal month/year &14.59 & 14.94 &  15.13 & 15.54 & 15.96 &
17.35 & 20.23 & 21.99 & 26.49\\ \hline solar days/year & 435 & 456
& 467.9 & 493.2 & 520.3 & 615.4 & 835.9 & 988.6 & 1434\\ \hline
length of solar day (hr)  &20.1 & 19.2 & 18.7 & 17.7 & 16.8 & 14.2
& 10.5 & 8.8 & 6.1\\ \hline
\end{tabular}
\\
\vspace{1cm} \\ $*$ Time is measured in million years before
present
\end{table}
\end{document}